# Non-Stationary Spectral Matching by Unconstrained Optimization and Discrete Wavelet Transformation

Mohammadreza Mashayekhi[1]*, Mojtaba Harati[2]*, and Homayoon E. Estekanchi[1]

[1]Department of Civil Engineering, Sharif University of Technology, Tehran, Iran
[2]Department of Civil Engineering, University of Science and Culture, Rasht, Iran.

**Abstract**

This study presents a non-stationary spectral matching approach in which unconstrained optimization is employed to adjust the signal to match a target spectrum. Adjustment factors of discrete wavelet transform (DWT) coefficients associated with the signals are considered as decision variables and the Levenberg-Marquardt algorithm is employed to find the optimum values of DWT coefficients. The proposed method turns out to be quite effective in the spectral matching objective, where matching at multiple damping ratios can be readily achieved by applying the proposed method. The efficiency of the procedure is investigated in a case study and compared with two conventional spectral matching methods. Results show considerable improvement in the matching accuracy with minimal changes in shaking characteristics of the original signal.

## 1. Introduction

Depending on the seismicity of the site and characteristics of the structure, various analysis methods can be considered for seismic assessment of a new or existing building. There are several conditions where dynamic time history analysis cannot be avoided, i.e. for complex structures equipped with base isolators, viscous dampers, and similar devices. In the dynamic time history analysis, it is required to select a set of ground motions that have the same seismicity characteristics expected at the site. The seismicity of the site is usually represented by a uniform hazard spectrum, which is normally prescribed by the codes using a design target spectrum. Since there are not adequate naturally recorded ground motions for many places, artificial or modified ground motions may be decided to be generated in a way that their response spectra would be compatible with a target spectrum (Bommer and Acevedo 2004, Iervolino et al. 2010).

Artificial spectrum-compatible ground motions, which are available from the literature (e.g. Rodolfo Saragoni and Hart 1973, Boore (1983, 2003), Atkinson and Silva 2000), may be produced by different methods considering some factors such as predefined target response spectrum, the fault-to-site distance, earthquake magnitude and characteristics of the fault and site as well. While significant advancements have been made for generating of such artificial ground motions, they are not always preferred for use in the time history analysis because artificially generated motions may differ somewhat from the real ground motions when they are compared for some characteristics, including the number of cycles, frequency content, duration of motion and etc. (Galasso et al. 2013). On the other hand, and in the process of producing modified earthquake records, the frequency content of the original acceleration time series is adjusted using spectral matching methods in such a way that their response spectra get matched to a target spectrum at all or a range of considered natural periods of vibration. The non-stationary characteristics, as well as the energy and the frequency content of the original earthquake, are almost preserved for the matched or adjusted records although the earthquake time series are also manipulated in this way (Lancieri et al. 2018).

---
* These authors have contributed equally to the work
Corresponding: Mohammadreza Mashayekhi, Research Associate, Sharif University of Technology, Tehran, Iran.
Email: mmashayekhi67@gmail.com

Engineers can also use the scaling method to circumvent the problems associated with the two methods mentioned above: the adjusted and artificial spectrum-compatible ground motions. For obtaining a scaled ground motion, the original acceleration time series of the earthquake is multiplied by a constant factor so that its response spectrum equals or exceeds the target spectrum in a vicinity of a particular structural natural period (Iervolino and Cornell 2005, Bommer and Acevedo 2004). Due to the inherent variability of the natural ground motions, more dynamic analyses may be required for the seismic response of the structures when scaled records instead of matched motions are intended to be employed. Using scaled records may lead to large structural dispersions in the dynamic analysis (Jayaram et al. 2011), which can be time-consuming and expensive in case complex structures are under investigation. However, matched or adjusted versions of real earthquake ground motions can be employed instead in order to substantially reduce the computational time related to such dynamic analyses (Bazzurro and Luco 2006, Watson-Lamprey and Abrahamson 2006, Carballo 2000, Shahryari et al. 2019). Another byproduct of using spectral matching is that any real earthquake record can be selected to produce a matched motion regardless of the strict record selection criteria prescribed by the codes for a specific area. In this case, engineers would have much more choices for ground motion selection, and they can have no concern for not having relevant expertise knowledge to select appropriate ground motions by themselves. Then, using spectral matching seems to be a better choice than the scaled motions or the ones generated artificially.

Matched ground motions are also utilized in time history analyses of the studies being accomplished to the find potential impacts of motion duration on structural responses. In this field of study, researches (e.g. Bravo-Haro and Elghazouli 2018; Hancock and Bommer 2007; Xua et al. 2018) use spectral matching techniques to remove or diminish the influences of amplitude-based intensity parameters of the motions—the spectral acceleration for example—on the computed structural seismic demands. In this way, the matched ground motions which are employed in such studies only differ in terms of duration and non-stationary characteristics (Hancock and Bommer 2007). Consequently, we can readily detect the correlational behavior or sensitivity of induced damages of a specific structure (or structural system type) to the motion duration of the selected ground motions.

The spectral matching can be done within two domains, namely the frequency and time domains (Preumont 1984). Spectral matching in the frequency domain is accomplished by modifying the Fourier amplitude spectra of the real acceleration time series (Rizzo et al. 1975, Silva and Lee 1987). Although this method is so simple and uncomplicated, it was reported that it doesn't have a good convergence status (Atik and Abrahamson 2010). Also, the adjusted motion is altered in such large extends that an offset appears at the end of its velocity and displacement profiles, so it urgently needs a baseline correction post-processing (Shahbazian and Pezeshk 2010). Besides, the frequency-domain method leads to large changes in the non-stationary characteristics of the acceleration, velocity, and displacement waveforms. Accordingly, the adjusted motion would become quite dissimilar from a real earthquake time history and changed into a motion having an unrealistic high energy (Naeim and Lew 1995).

In the time-domain spectral matching, wavelets are used and added to the acceleration time series, introducing less energy into adjusted ground motions though it causes the problem to become highly nonlinear (Adekristi and Eatherton 2016). The first algorithm, in this case, was offered by Kaul (1978) and then extended by Lilhanand and Tseng (1988) to incorporate a new mother wavelet that ensures numerical stability and convergence. However, the extended algorithm failed to maintain the non-stationary characteristics of the original motion. By adopting a new adjustment wavelet, Abrahamson (1992) developed RspMatch software by implementing the algorithm of Lilhanand and Tseng (1988). Although the updated algorithm could well preserve the non-stationary characteristics of the real ground motions, drifts were yet observed in the resulted velocity and displacement histories of the matched records. Therefore, they required a baseline correction after the matching process. Subsequently, Hancock et al. (2006) revised RspMatch program using combinational adjustment wavelets, composing of the functions proposed by Abrahamson (1992) and Suarez and Montejo (2003, 2005), to further address nonlinearity of the problem and include baseline correction in the functional form of the algorithm. This revised algorithm also lets us match the ground motion records to pseudo-acceleration target spectra with different damping levels at the same time. While fewer changes are observed in the final matched records filtered using the method proposed by Hancock et al. (2006), more wavelets may be sometimes essential to get added to the motions for the purpose of improving the convergence status of the numerical solution. Then Atik and Abrahamson (2010) proposed an adjustment function, namely the improved tapered cosine wavelet basis, which is replaced with the wavelets used by Hancock et al. (2006) in the RspMatch program. As a result, this wavelet replacement guarantees a stable and fast matching algorithm, a better convergence status, and an analytical solution without any need for a baseline correction post-process. Recently, Adekristi and Eatherton (2016) offered the Broyden Updating to solve the nonlinear equations of spectral matching and reported that the new proposed procedure is successful in maintaining the main seismic characteristics of the original motions in the matched records.

Optimization methods are a very robust technique, which can be easily implemented and applied to many engineering problems. Even though the current time-domain spectral matching methods set the amplitude and timing of the adjustment wavelets utilizing a least-squares formulation, optimization algorithms have not been widely employed in the methods for spectral matching of the ground motions. For example, the amplitudes of the adjustment wavelets in the current methods are in fact set by minimizing the least-squares, which works based on a linearization of the problem at each iteration. However, Alexander et al. (2014) introduced an optimization-based algorithm for changing any accelerogram to be matched with the desired target spectrum. They employed the Voltera series for expressing time series and Levemberg-Marquardt algorithm for finding the optimal time series. In the Volterra series representation of a signal, it is required to adopt appropriate Kernel functions for signal representations. Alexander et al. (2014) estimated first-order Volterra Kernel by multi-level wavelet decomposition. Higher-order Volterra Kernel is then estimated by multi-modal mixing of the first-order kernel functions. Their results showed a robust fit to the spectrum. Besides, Jayaram et al. (2011) proposed a novel ground motion selection procedure that is equipped with a technique called greedy optimization method, which can be opted to efficiently collect earthquake records that are fairly matched to a target response spectrum mean and variance. Additionally, optimization methods have been also hired to generate endurance time (ET) excitations (see: Mashayekhi et al. 2018a, Mashayekhi et al. 2018b). In this case, Mashayekhi et al. (2018c) combined wavelet transform with an optimization method to produce ET spectrum-compatible ground motions. It is worth to be mentioned that yet another research by Nakhaeim and Mohraz (2010) also utilized the wavelet transformations to investigate the inelastic spectral matching.

In this paper, an optimization-based approach is proposed through which an optimization algorithm is used for the spectral matching in a way that discrete wavelet transformation (DWT) coefficients of a signal are modified for making a ground motion be compatible with the predefined target spectra. The adjustment factors of DWT coefficients are selected as decision variables in this developed optimization procedure, and the Levenberg-Marquardt algorithm is then used to find the optimum values of the decision variables. The algorithm is devised in such that while the non-stationary characteristics of the motions are finely preserved, matched motions readily integrate to zero velocity and displacement without any extra post-processing procedure such as baseline correction. It also provides a stable solution without introducing drift to the final velocity and displacement histories of the adjusted motion at the end, which can be simultaneously applied to target pseudo-acceleration response spectra with multiple damping levels.

## 2. Theoretical basis

### 2.1. Wavelet decomposition

Fourier Transform (FT) and Fast Fourier Transform (FFT) are efficient tools for finding frequency components of signals, but they are not suitable and applicable for non-stationary signals such as the ones representing ground motion records. The reason is that in FT and FFT, frequency components are obtained from an average over the whole length of signals and hence frequency changes over time cannot be captured. In contrast to FFT, wavelet analysis characterizes local features of signals and is a powerful tool for time-frequency analysis. Therefore, wavelet analysis can capture frequency changes in signals and is applicable for earthquake non-stationary motions.

The purpose of the classic wavelet is to break a signal down to its constituent parts. Unlike FFT that always uses sinusoidal functions to decompose a signal, wavelet analysis uses translated and scaled wavelet functions. Therefore, two parameters—namely the scale and translation—are needed in order to define the wavelet functions. Different scales and transitions, respectively, produce different frequencies and times in the analysis. Wavelet functions are represented as follow:

$$\psi_{a,b} = \frac{1}{\sqrt{a}} \psi\left(\frac{t-b}{a}\right) \quad a,b \in \mathbb{R}, a \neq 0 \qquad (1)$$

where $a$ and $b$ are scaling and translation parameters controlling frequency and time in wavelet analysis, where $\psi$ is a function called the mother wavelet. The properties of mother wavelet can directly influence the properties associated with the basis functions. The term "mother" implies that the functions that are used in wavelet analysis are derived from one main function or the mother wavelet. This study used db12 (Daubchies12) as the mother wavelet (Daubechies, 1992).

Continuous Wavelet Transform (CWT) is expressed as follow:

$$W_\psi f(a,b) = \frac{1}{|a|^{1/2}} \int_{-\infty}^{\infty} f(t) \bar{\psi}_{a,b}(t) dt \quad (2)$$

where $W_\psi f(a,b)$ is a wavelet transform coefficient that represents how well signal $f(t)$ and wavelet functions match. And $\bar{\psi}_{a,b}(t)$ is a complex and conjugate of the wavelet functions. So, the signal $f(t)$ can be reconstructed by using Equation (3),

$$f(t) = \frac{1}{2\pi C_\psi} \int_{-\infty}^{\infty} \int_{-\infty}^{\infty} \frac{1}{a^2} W_\psi f(a,b) \psi_{a,b}(t) \mathrm{d}a \mathrm{d}b \quad (3)$$

where $C_\psi$ is

$$C_\psi = \int_{-\infty}^{\infty} \frac{|\hat{\psi}(\omega)|^2}{|\omega|} \mathrm{d}\omega < \infty \quad (4)$$

And $\hat{\psi}(\omega)$ is the Fourier transform of $\psi(t)$ at the frequency of $\omega$

The integral of Equation (3) has to be discretized for numerical evaluation. Discrete Wavelet Transform (DWT) is similar to CWT, except for the possible choices we have for ($a$, $b$). There is no constraint for the choice of (a, b) in the CWT, however, these parameter values in DWT are restricted as follow:

$$a = a_0^j, \quad b = kb_0 a_0^j \quad (5)$$

where $j$ and $k$ are members within the set of all positive and negative integers. In addition, this study focuses on orthogonal wavelet bases and chooses $a_0 = 2$ and $b_0 = 1$, respectively.

The DWT can be implemented as a set of filter banks—comprising a high-pass and a low-pass filter—each of which followed by a down-sampling procedure which converts them into two components. Decomposition of a signal, splitting it into high and low resolutions, is only applied to low pass channel as shown in Figure 1 (a). In fact, the output of the low-pass filter is also passed through low-pass and high-pass filters. This process continues until the desired decomposition level. As can be seen from Figure 1 (a), the decomposition procedure is performed at three levels. It should be mentioned that the downward arrow enclosed within a circle, as can be seen in Figure 1 (b), is a symbol standing for the down-sampling function. The down-sampling is actually an operator through which every second data point is removed from the signal. Likewise, low-pass and high-pass filters are also called by scaling and wavelet filters, respectively. Spectral density of high-pass and low-pass filters of Daubechies12 versus angular frequency is demonstrated in Figure 2. These filters—which are in contrast to pass-band filters whose amplitudes are one at pass-band regions—are designed to determine the discrete wavelet coefficients rather than filtering the signals only.

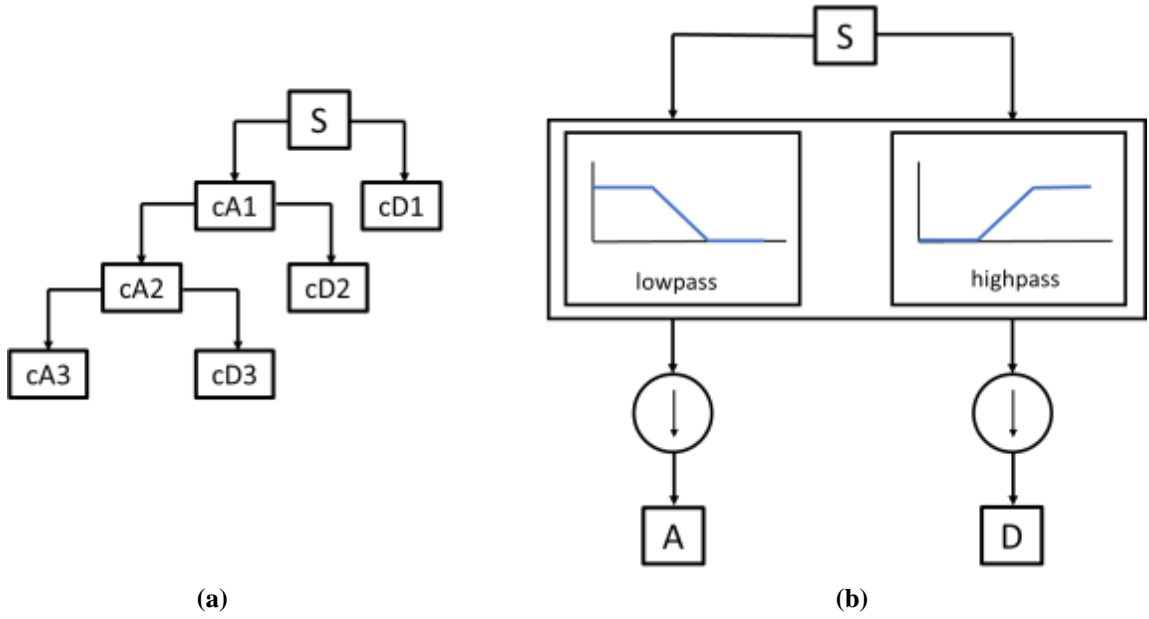

**(a)**  **(b)**

**Figure 1 Tree decomposition of the DWT procedure**

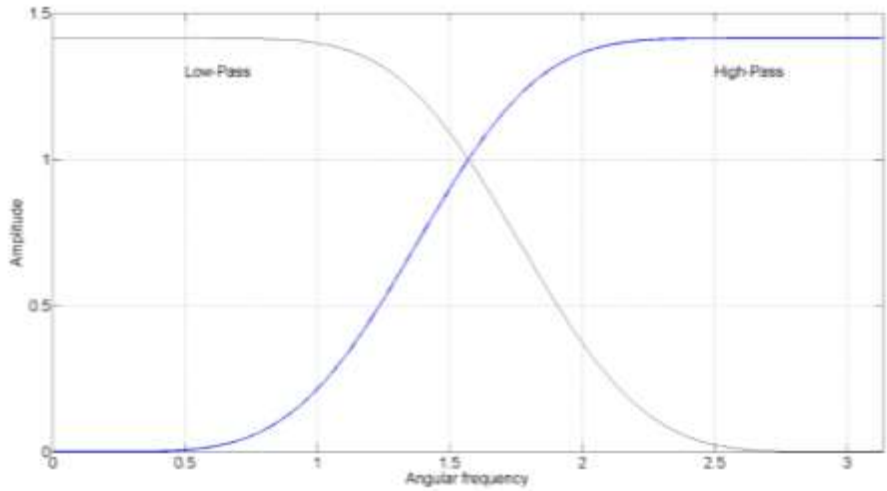

**Figure 2 Power spectral density for high-pass and low-pass filters of Daubechies 12**

The process for assembling back the components of a signal into the original signal is called reconstruction or synthesis. Wavelet analysis involves filtering and down-sampling while the wavelet reconstruction process consists of up-sampling and filtering. Up-sampling is a process of lengthening a signal component by inserting zeros between samples. The general process for such reconstruction is depicted in Figure 3. It should be noted that the encircled upward arrow, as shown in Figure 3, is a symbol for up-sampling. If all DWT coefficients set to zero—except for cD1—the reconstructed signal would be equal to the detailed form of a signal at level 1 and is denoted by D1. Similarly, if all DWT coefficients are set to zero, except for the cD2, the reconstructed signal is the detailed form of the signal S at level 2 and is denoted by D2. These reconstructed signals ($D_1$, $D_2$) are also known as DWT components of a signal. In this case, a signal can be reconstructed by its DWT components as expressed by Equation (6).

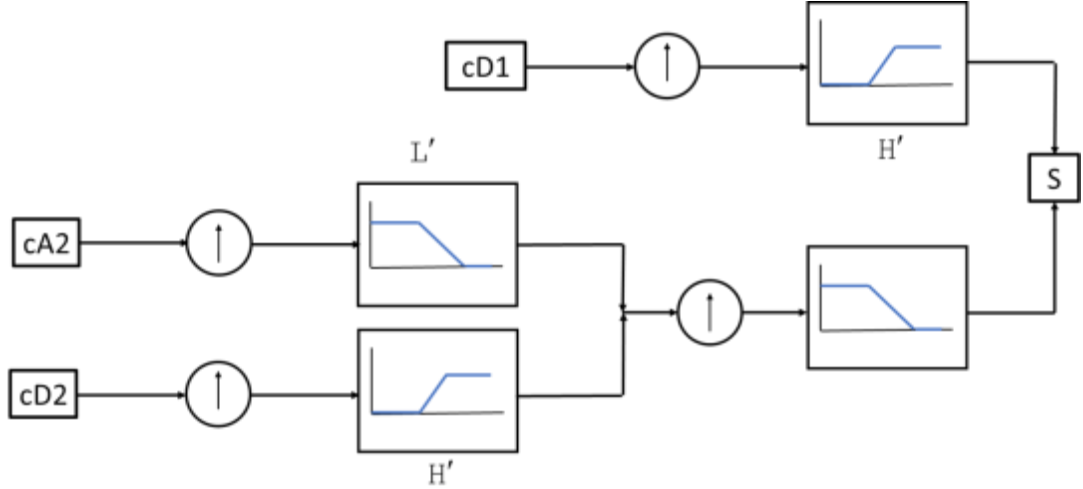

**Figure 3 Reconstruction process of a signal in the DWT framework**

$$S = A_J + \sum_{j=1}^{J} D_j \qquad (6)$$

where $A_J$ is the approximated form of a signal at level J, and $D_j$ is the detailed form of a signal S at level j. In this formula, the decomposition is carried out at J level. In this way, $A_j$ and $D_j$'s are DWT components of the signal S. It is worth mentioning that $A_J$ and $D_j$'s are functions on time and they are, in fact, time histories that are added together to reconstruct the signal S.

### 2.2. Nonlinear least square approach

The least-squares method, in general, is a problem of finding a vector *x* that is a local minimizer for a function that is a sum of squares as follow:

$$\min_x \|F(x)\|_2^2 = \min_x \sum_i F_i^2(x) \qquad (7)$$

The Levenberg-Marquardt (Levenberg 1994, Marquardt 1963) is an optimization algorithm which is used to find the local minimum of least square problem. Optimization algorithms, in general, start with an initial point. Each classic optimization algorithm uses a search direction in which a line search is performed to find the local minimum. This process continues with the determined local minimum at a previous iteration until convergence is achieved. The Levenberg-Marquardt algorithm uses a search direction between the Gauss-Newton direction and the steepest descent direction. This search direction is a solution of the linear set of equations as follow:

$$\left(J(x_k)^T J(x_k) + \lambda_k I\right) d_k = -J(x_k)^T F(x_k) \qquad (8)$$

where $J(x_k)$ is the Jacobian matrix of F at the point $x_k$, and $I$ is the identity matrix. The term $\lambda_k$ is a factor that controls both the magnitude and the direction of $d_k$. When $\lambda_k$ is set to zero, the direction $d_k$ would be identical to that of the Gauss-Newton method. As $\lambda_k$ tends toward infinity, $d_k$ approaches to the steepest descent direction coupled with a magnitude tending to zero. This implies that for some large values of $\lambda_k$, the term $F(x_k + d_k) < F(x_k)$ remains true. $\lambda_k$ can be therefore controlled to ensure a descending manner even if the second-order terms—which restrict the efficiency of the Gauss-Newton method—are encountered in this procedure. The major difficulty in the implementation of Levenberg-Marquardt is to find an effective strategy for controlling the size of $\lambda_k$ at each iteration.

The conditions for convergence criteria are given below. If each of these criteria is satisfied, the optimization process will stop and the local minimum of the final iteration will be the solution of the problem.

- Iteration number reaches a specified value. In this study, the maximum iteration number equal to 40 is considered.
- The number specified for function evaluations reaches a pre-defined value. In this research, the maximum number considered for function evaluations is equal to 400000.
- Size of the calculated step, which is the norm of $x_{k+1}$-$x_k$, goes below a specified value (
$$|x_k - x_{k+1}| < 10^{-8}(1 + |x_k|)$$
- Changes in the objective function value during a step are less than a predetermined value. Or equivalently: $|f(x)_k - f(x_{k+1})| < 10^{-8}$

## 3. Proposed method

The main purpose of this study is to adjust the DWT coefficients of a signal in order to be compatible with a specific target spectrum by using optimization methods. Two types of residuals are defined for quantifying the difference between response spectra of a signal and target response spectra considered. First, the residual function, as shown in Equation (9), computes absolute residuals while second residual function as brought in Equation (10) is based on the relative computation of the differences.

$$E(T,\xi) = S_a(T,\xi) - S_{aT}(T,\xi) \tag{9}$$

$$R(T,\xi) = (S_a(T,\xi) - S_{aT}(T,\xi))/S_a(T,\xi) \tag{10}$$

where $S_a(T,\xi)$ and $S_{aT}(T,\xi)$ are, respectively, acceleration spectra of a signal and the target motion at period $T$ with a damping ratio of $\xi$. It is worth to add that attempts—in this proposed method—are made to match a ground motion record to a selected acceleration spectrum for a range of damping values, so target spectra brought in Equations (9) and (10) are a function of the damping ratio hereafter.

This study uses nonlinear least-squares optimization to determine adjustment factors of DWT coefficient of a signal to get matched to the target acceleration spectra. In the optimization context, equations are expressed in term of objective functions. In this study, the objective function that is utilized to be at work is brought in Equation (11). As can be readily understood, this objective function is able to integrate absolute residuals over different periods and damping ratios.

$$F(S) = \int_0^1 \int_{T_{\min}}^{T_{\max}} (S_a(T,\xi) - S_{aT}(T,\xi)) \, dT \, d\xi \tag{11}$$

where $F(S)$ is the objective function value of a signal $S$, $T_{min}$ is the minimum considered period of vibration, and $T_{max}$ is the maximum considered period within the objective function. In order to evaluate this objective function, period and damping ratio have to be discretized. The type of discretization influences the final results. In this study, the period is sampled at 130 points: 120 points in the interval of [0.02sec, 5sec] that are logarithmically distributed while the rest 10 points would be within the interval of [5sec, 50sec]. Four cases for the damping ratio discretization are considered, where damping is only discretized at 5% level in case 1. In case 2, damping is discretized for values related to both 5% and 10% levels. In case 3, damping is discretized at 5%, 10%, and 20% simultaneously. Finally, values of 5%, 10%, 20%, and 30% are selected for discretizing damping ratio for case 4. To solve this objective function using nonlinear least-squares optimization, the objective function is expressed in a matrix format as follow:

$$[M_{ij}] = [S_a(T_i, \xi_j) - S_{aT}(T_i, \xi_j)] \quad 1 \leq i \leq n_T, 1 \leq j \leq n_\xi \quad (12)$$

where $T_i$ and $\xi_j$ are the discrete values of period and damping ratios. $n_T$ and $n_\xi$ are the number of discretization points for period and damping ratio. In this study, $n_T$ is 130 and $n_\xi$ is 1, 2, 3 and 4 for cases 1, 2, 3 and 4, respectively.

In addition to an objective function, decision variables have to be specified in the optimization context. In the proposed spectral matching, decision variables are adjustment factors of DWT coefficients of a signal. For example, detailing coefficients of the adjusted signal is determined through Equation (13). In this equation, $cD_1$ and $cD_1^{adjust}$ are the level 1 detailing coefficients of the original and adjusted signal, respectively. $x_{D1}$ is an optimization variable associated with level 1 detailing coefficients and has the same dimension of $cD_1$. Each component of this variable changes the DWT coefficients of the adjusted signal.

$$cD_1^{adjust} = x_{D1} \times cD_1 \quad (13)$$

Although $x_{D1}$ may be considered to become an optimization variable that is associated with detailing coefficients of level 1, adjustment factors of different decomposition levels can also be taken as decision variables. In this case, total decision variables in the optimization procedure of the proposed non-stationary spectral matching depend on the number of considered decomposition levels. The more levels are considered in this case, the more optimization variables are created. Large numbers of optimization variables need more time for convergence of the final solution, but it may lead to a more accurate adjusted signal. Here the accuracy is quantified by matching degree acquired against the target response spectra. Regarding the accuracy and expected computational time simultaneously, adjustment factors of approximation coefficients of level 9 and detailing coefficients of levels 1-7 are considered as decision variables. In fact, adjustment factors for detailing coefficients of levels 8 and 9 are not considered as the decision variables. Because several trial and error evaluations demonstrate that the adjustment factors of levels 8 and 9 have a negligible impact (less than 1%) on the accuracy, but they severely increase the computational time.

An optimization process starts with initial values for decision variables. For initializing optimization process of this study, all starting adjustment factors are taken to be 1. The reason is that it is desired to achieve the best match while overall shaking characteristics of the signal are not extensively changed. Another important matter in the non-stationary spectral matching is that the acceleration, velocity and displacement time histories should get to zero at the end of these signals. Because there is no control over this property in the unconstrained optimization process, the velocity and displacement at the end of a matched motion may become nonzero. In order to resolve this problem, two wavelet-form sine functions are added into the signal to impose this condition at the end of adjusted motions. This technique called a primitive baseline correction hereafter is schematically displayed in Figure 4. The adjusted signal with the primitive baseline correction is also computed according to Equation (14).

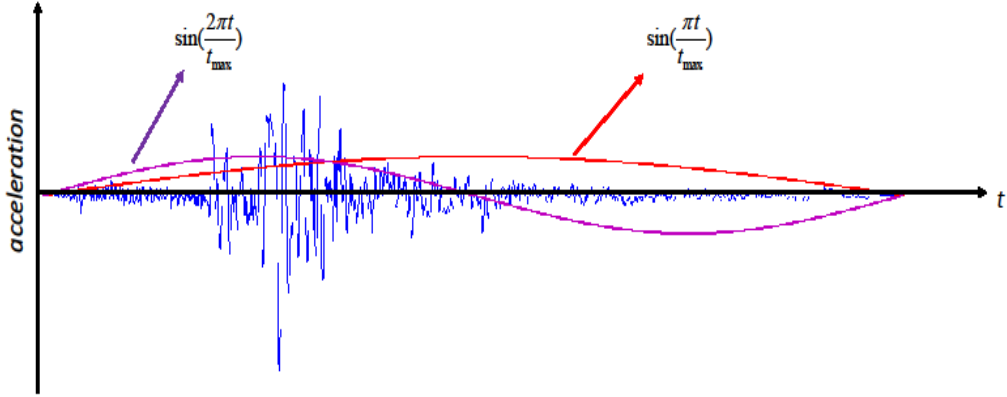

**Figure 4 Schematic view of the proposed method for a primitive baseline correction**

$$a^{blc}(t) = a^{orig}(t) + \alpha \sin\left(\frac{\pi t}{t_{end}}\right) + \beta \sin\left(\frac{2\pi t}{t_{end}}\right) \qquad (14)$$

where $t_{end}$ is the duration of the signal, $a^{blc}$ and $a^{orig}$ are primitively baseline-corrected and original signals, respectively. α and β are contribution factors of two sine functions in the primitively baseline-corrected signal. The coefficients of α and β are calculated from velocity and displacement at the end of an original signal. These coefficients are calculated as follow:

$$\alpha = \frac{-\pi \times v_{t_{end}}}{2 t_{end}} \qquad (15)$$

$$\beta = \frac{-2\pi \times x_{t_{end}}}{t_{end}^2} - 2\alpha \qquad (16)$$

where $v_{t_{end}}$ and $x_{t_{end}}$ is the velocity and displacement at the end of an original signal. It should be also mentioned that the value of α and β are so infinitesimal (on the order of $10^{-4}$), so it can be expected that such superimposed functions do not have any strong effect on the characteristics of the original signal.

The initial or primitive baseline correction which is based on two sine functions would probably work well for the end drifts, but it is not anticipated to be adequate for the time series employed in the engineering applications because the baseline drift is not only related to the ending velocity and displacement values. For the ground motions used in engineering projects, the urgent concern is the long period drift throughout the time history. In this case, the standard methods of baseline correction fit the uncorrected displacement to a high-order polynomial (typically on the order 5). Therefore, an extra baseline correction, developed by Boore (1999), has been already implemented and placed at the end of the proposed algorithm for spectral matching. The summary of the proposed algorithm is presented in Figure 5.

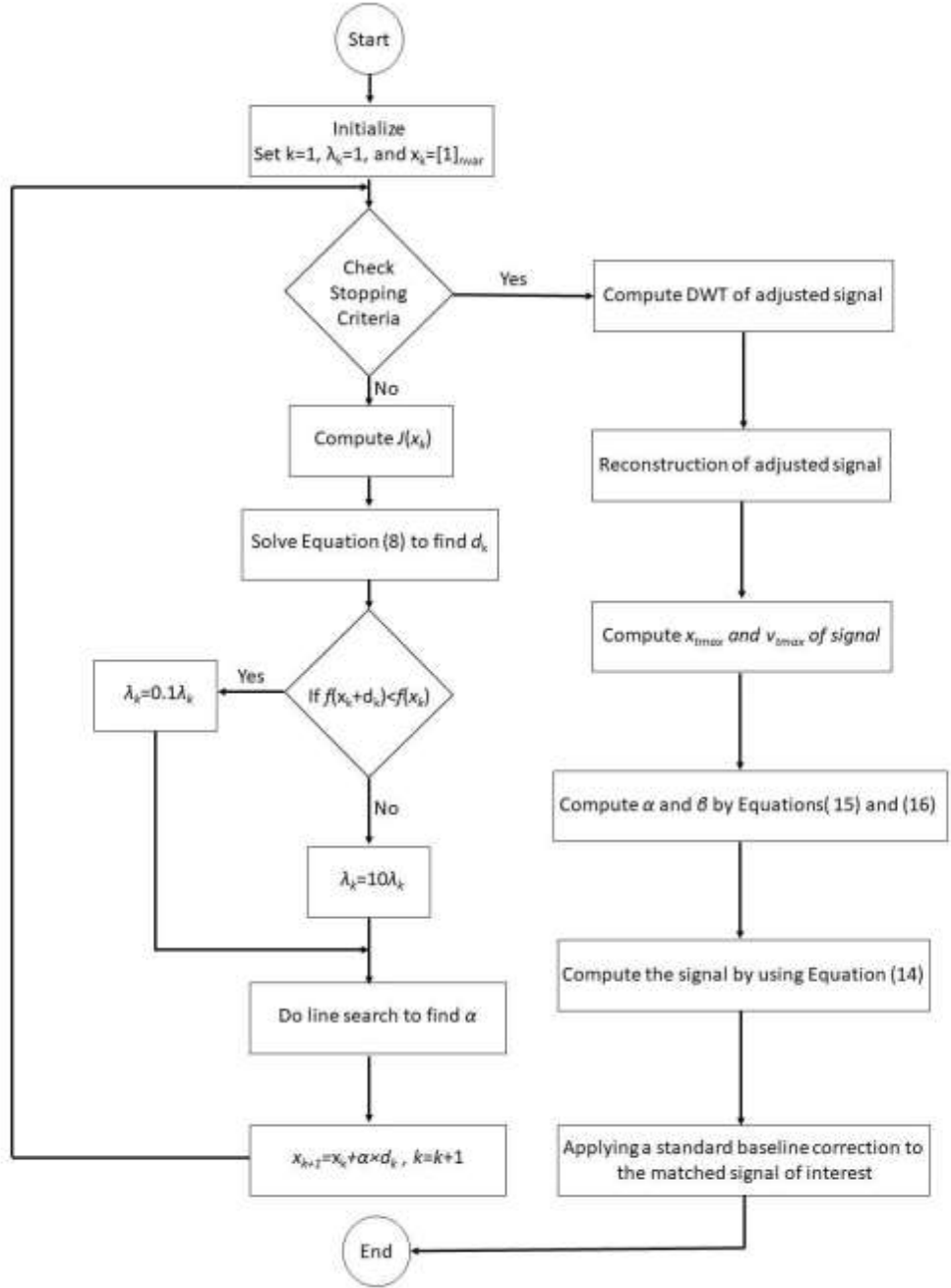

**Figure 5 The proposed algorithm for non-stationary spectral matching**

In order to quantify the expected misfit of the adjusted signal, Equation (17) is proposed. In this equation, the average misfit of the signal with target acceleration spectra at each damping ratio is expressed in percent.

$$M\left(\xi\right) = \frac{\int_{T_{\min}}^{T_{\max}} \left(S_a\left(T,\xi\right) - S_{aT}\left(T,\xi\right)\right) \mathrm{d}T}{T_{\max} - T_{\min}} \tag{17}$$

## 4. Numerical example

To show the capability of the proposed method to produce a spectrum-matched accelerogram, it is applied to the record of 1989 Loma- Prieta earthquake in order to make it compatible with predefined target acceleration spectra. The selected motion was recorded at a station, i.e. the Diamond Heights (record 00794T in the NGA database), located 71 km from the fault rupture. The target acceleration spectra and the seed accelerogram—namely the Loma- Prieta earthquake of 1989—are the same as the ones opted by Hancock et al. (2006). They have been selected in this way because the results found by two approaches, outputs by the proposed method and the ones through Hancock et al. (2006), would become readily comparable with each other. It should be also noted that the above-mentioned ground motion is linearly scaled by a factor of 3.274, both in this study and the one accomplished by Hancock et al. (2006). This section of the paper incorporates three subsections, including the matching procedure regarding a single damping ratio spectrum and the spectral matching for multiple damping levels. First, the procedure and the results associated with fitting the selected seed accelerogram to the 5% damped target response spectra are discussed. Then it is shown how this method can be effectively applied to generate an adjusted motion that matches multiple damping levels at the same time.

### 4.1. Matching to single damping ratio spectra

The linearly scaled ground motion is adjusted with the proposed optimization-based spectral matching so that it matches target acceleration spectra for a set of periods defined from 0.05-5 sec. Figure 6 (a) compares the acceleration spectra of adjusted and original motion with the 5% damped target acceleration spectrum for periods ranging from 0.05 to 5 sec, where they are respectively shown in green, red and blue color. As can be seen, there is a close match between the response spectrum of adjusted motion and the target spectrum. The average misfit for adjusted accelerogram, which is computed from Equation (17), is 0.4% while this quantity was about 20.3% for linearly scaled (unmatched) accelerogram. This is a reliable indicator of how effectively the proposed method is able to enhance the goodness-of-fit to the target spectrum. Comparison between displacement spectra of the adjusted and original motion and the response spectrum associated with the target spectral displacement are also provided in Figure 6 (b). It is apparently realized that adjusted motion is neatly fitted to the target displacement spectrum as well.

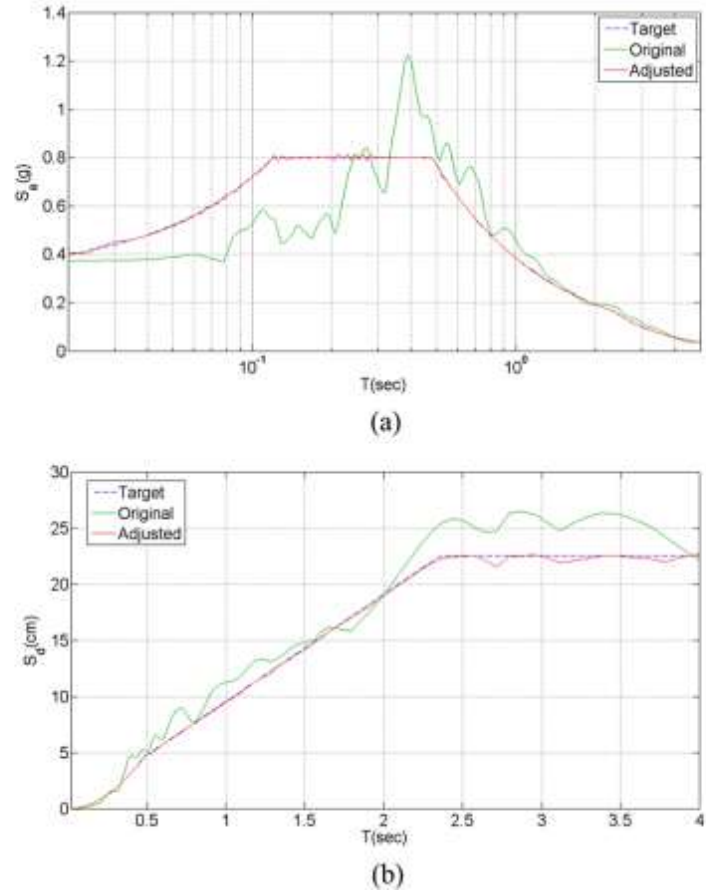

**Figure 6 Comparison of response spectra of adjusted accelerogram, target, and original accelerogram: (a) acceleration spectra; (b) displacement spectra.**

Acceleration, velocity, and displacement time histories of adjusted accelerogram, shown in green color, are presented and compared with the profiles of the original ground motion which are shown with blue color as demonstrated in Figure 7 (a) to (c). It is quite interesting to observe that the proposed method is not only capable of showing a pronounced fitness to the target spectra but also able to well preserve the non-stationary characteristics of the original ground motion within the adjusted motion. While the amplitudes of matched velocity and displacement profiles are different from the ones pertinent to the original motion, adjusted records inherit the same general characteristics from the initial time series. Also, it is worthy of mention that there is no sign of end drifts in the acceleration, velocity, and displacement of the new computed time series. Moreover, by checking the ground motion profiles before and after the adjustment procedure, it can be understood that all the matched time series terminates to zero. It is reasonably consistent with the physical concept of a real earthquake at the end of a motion.

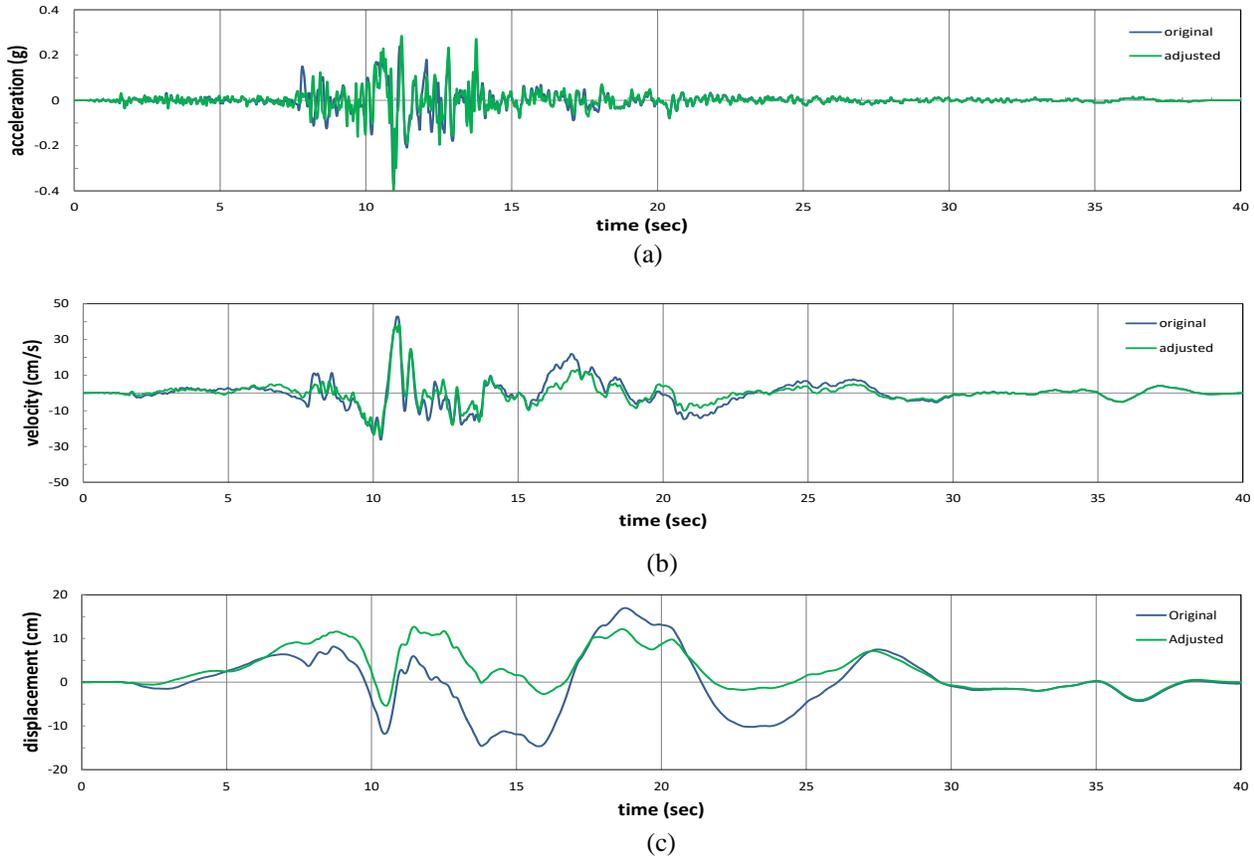

**Figure 7 (a) Acceleration time history of adjusted accelerogram vs. the original; (b) velocity time history of adjusted vs. original accelerogram; (c) displacement time history of adjusted vs. original accelerogram**

Furthermore, the build-up Arias intensity of adjusted accelerogram is compared with its counterpart for the original accelerogram in Figure 8. This demonstrates that the energy distribution of the adjusted and original motion is rather similar to each other and the difference between Arias intensity of adjusted accelerogram and original motion is less than 10%. Hence, this can show how efficacious this method would be in producing adjusted ground motions by retaining the main non-stationary characteristics as well as the energy content of the original motion.

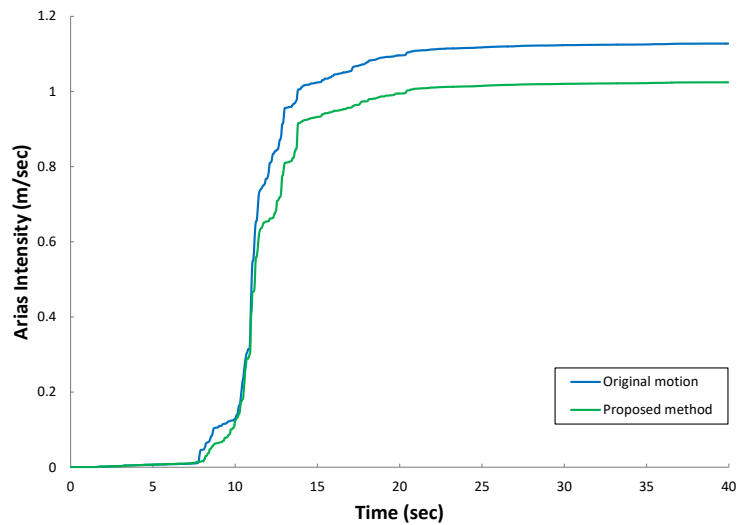

**Figure 8 Comparison of the Arias intensity in the adjusted accelerogram and original accelerogram**

**4.2. Matching to multiple damping levels**

In this section, the proposed spectral matching technique is applied to the initial motion in a way that it is adjusted to get matched at several damping levels. The matching of response spectra—which are associated with the adjusted motion—to the targets occurs within specified periods from 0.05 to 5 sec. In case 4 that was previously defined (in section 3), four damping ratios are simultaneously considered, namely 5%, 10%, 20%, and 30%. Figure 9 compares the acceleration spectra of adjusted and the original accelerogram at four damping ratios, where a comparison on relevant displacement spectra is also provided. As can be witnessed, there are reasonably good matches between the acceleration spectra of the adjusted time series and the targets at different damping ratios.

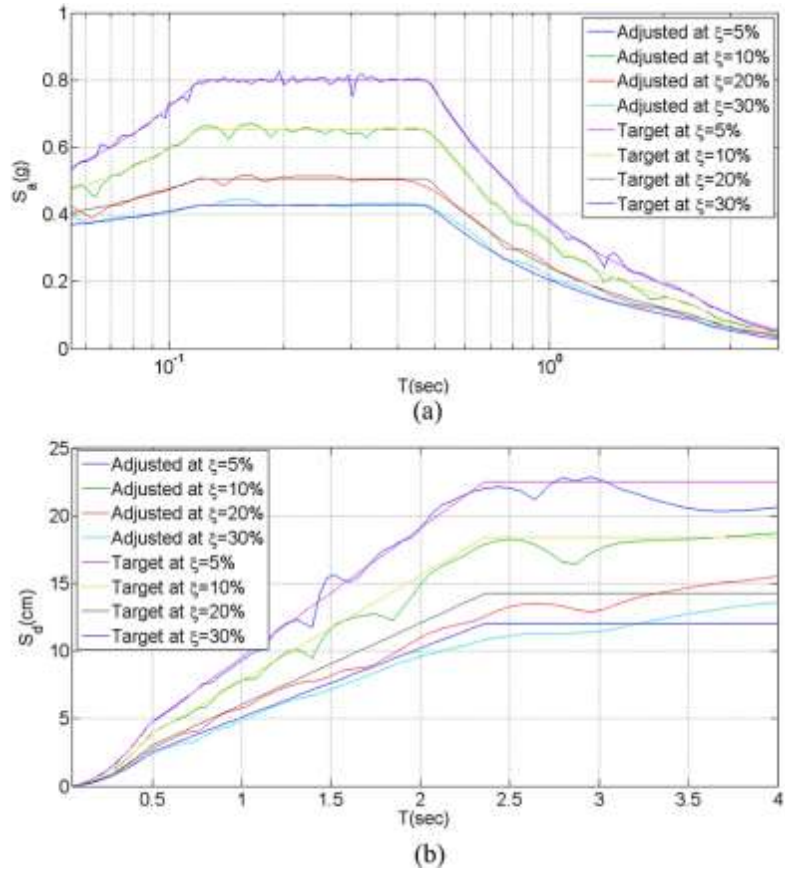

**Figure 9 Comparison of response spectra of adjusted accelerogram—at 5%, 10%, 20%, and 30% damping ratio—with the targets for: (a) acceleration spectra, and (b) displacement spectra.**

Table 1 summarizes the spectral misfit of original accelerogram and the ground motion adjusted for different damping ratios. As it is reflected in the table, the average misfit values pertinent to the case 4 are reduced by a factor of 7 when it is compared with the corresponding misfit values for the original (initial) motion considering different damping ratios. The optimization-based spectral matching approach is prosperous to reduce the associated amounts of misfit, thereby improving the quality of goodness-to-fit of the matched spectra in case multiple damping levels are encountered.

**Table 1 Average spectral misfit of adjusted accelerograms at different damping ratios**

| Damping level matched | Damping level | | | | |
|---|---|---|---|---|---|
| | 5% | 10% | 20% | 30% | All |
| Original | 20.3 | 18.4 | 21.8 | 31.6 | 23.5 |
| 5% | 0.4 | 6.3 | 18.0 | 29.0 | 13.4 |
| 5%, 10% | 1.0 | 1.9 | 10.6 | 20.5 | 8.5 |
| 5%, 10% 20% | 1.1 | 1.8 | 2.1 | 8.2 | 3.3 |
| 5%, 10%, 20%, 30% | 1.7 | 1.8 | 2.5 | 4.1 | 2.5 |

Acceleration, velocity, and displacement time histories of adjusted and original accelerograms are demonstrated in Figure 10. The original and the adjusted time series are represented by a blue and green line, respectively. It can be seen that the general non-stationary characteristics of the original motion have been appropriately conveyed to the adjusted time series though the amplitudes of the velocity and displacement profile of new computed time series are somewhat different to those observed from original motion. It should be noted that there is no sign of drifts again in the acceleration, velocity, and displacement of the adjusted time series at the end of the motion. While the adjusted time series integrates to zero at the end of the motion and consider the fact that a standard baseline correction has been already implemented in the proposed method, there is no need for a separate baseline correction after the matching process.

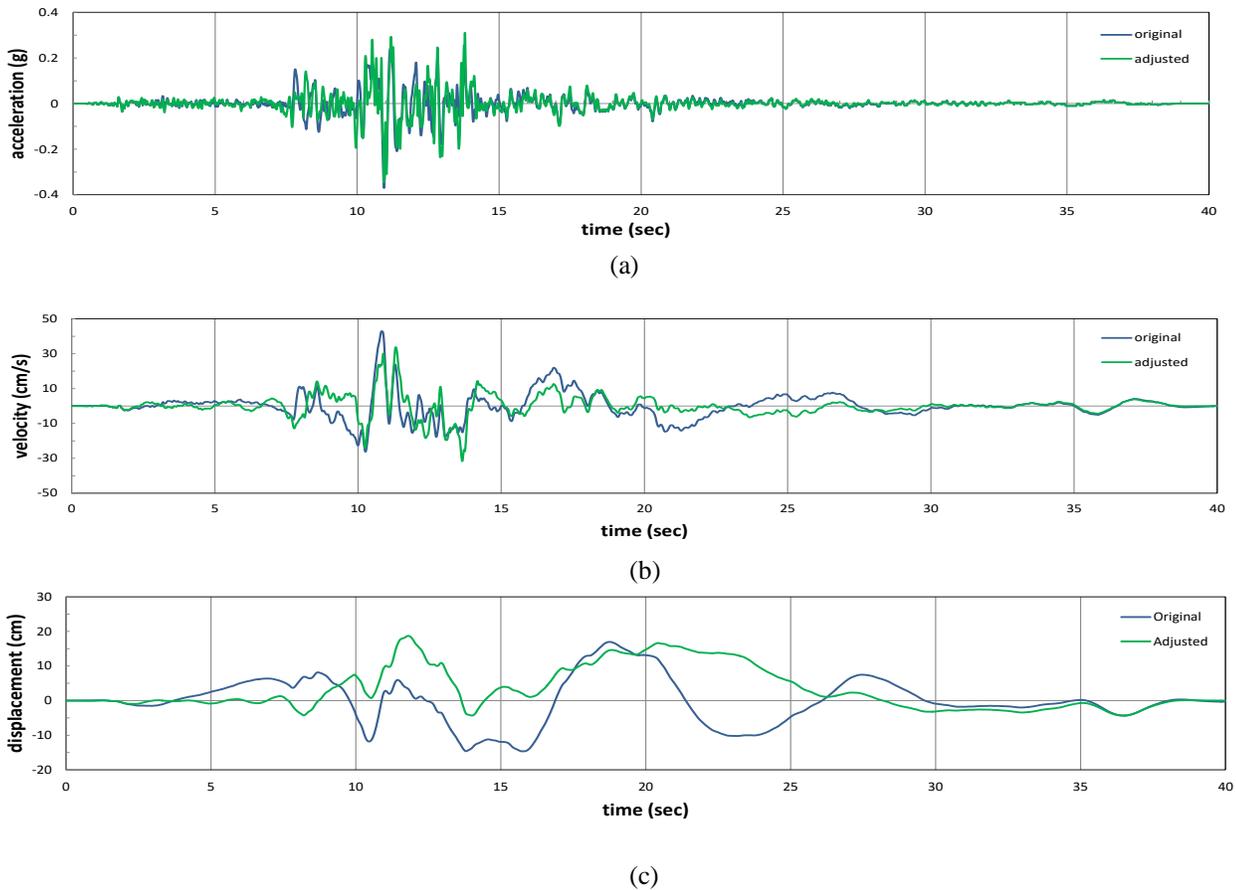

(a)

(b)

(c)

**Figure 10 Time histories of adjusted motions for 5%, 10%, 20%, and 30% damping ratios against the original ground motion: (a) acceleration time series (b) velocity profile, and (c) displacement record**

The Fourier amplitude and the build-up of Arias intensity are demonstrated for the original and adjusted accelerograms in Figures 11 and 12. We see that the Fourier amplitude of the adjusted time series retains somewhat similar characteristics of the original motion. And we can also find out that they share almost the same trends in their Fourier amplitudes. The difference between the Arias intensity of two motions—namely the adjusted and original ones—is about 1% here that means a decrease of 9% as compared to the adjusted accelerogram obtained considering a target spectrum with a 5% damping ratio only.

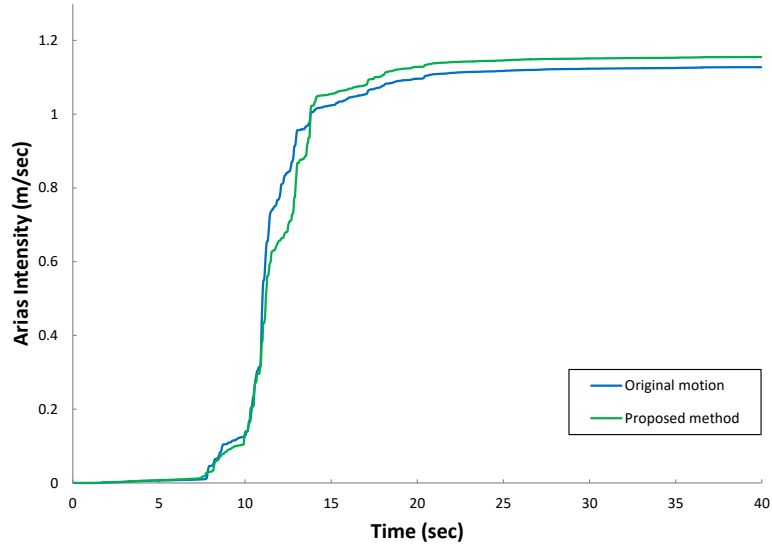

**Figure 11 Arias intensity of original accelerogram versus the earthquake record adjusted for 5%, 10%, 20%, and 30% damping ratios**

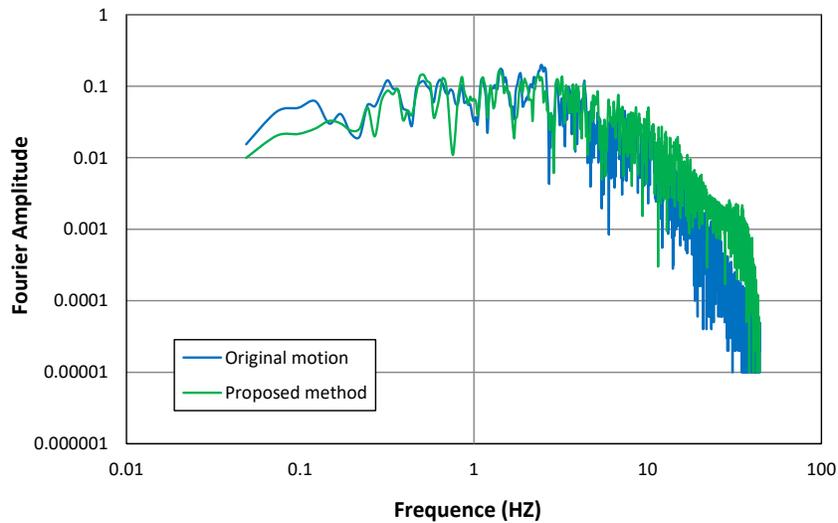

**Figure 12 Fourier amplitudes of the original accelerogram record (for a 5% damping ratio) and the adjusted motion simultaneously matched for different damping ratios—a 5%, 10%, 20% and 30% damping factor**

## 5. Comparison with existing methods

The results obtained by the proposed method are compared with the outcomes of two existing robust approaches, including the methods developed by Hancock et al. (2006) and Atik and Abrahamson (2010). The outcomes of two existing approaches are generated using the SeismoMatch (2016). This is a program containing the codes presented by two aforementioned methods. The spectral matching procedures are conducted—utilizing the proposed method as well as the ones mentioned above—for a range of periods defined from 0.05 to 5 sec. In this case, the 5% damped target response spectrum and the seed accelerogram are the same as the ones employed in the previous section of the paper. Figure 13 compares the acceleration spectra of the original motion and adjusted time series produced by different methods. Although the proposed matching procedure is developed using a different method completely, we can see that it produces similar outcomes as compared with the other existing techniques. Besides, the average misfit value obtained for the method proposed is also improved and become 0.4% while it is 1% and 2.1% for the approaches developed by Hancock et al. (2006) and Atik and Abrahamson (2010), respectively.

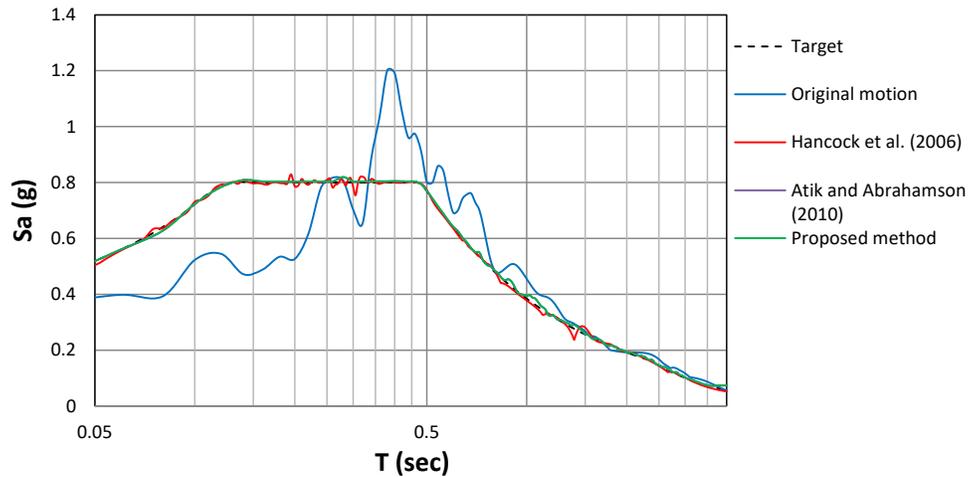

**Figure 13 Comparison of response spectra of the original accelerogram against adjusted motions using different spectral matching procedures**

As can be seen from Figure 14, a comparison has been made on the build-up of normalized Arias intensity of the adjusted ground motions generated with different spectral matching procedures. We see that all the selected spectral matching procedures share the same general energy distribution patterns which are so similar to the one developed by a real initial ground motion. As a result, it is clear that the proposed method is able to produce a compatible-spectrum motion in a way that its energy content and the pattern of its general energy distribution follow the same trend of which the original ground motion is composed.

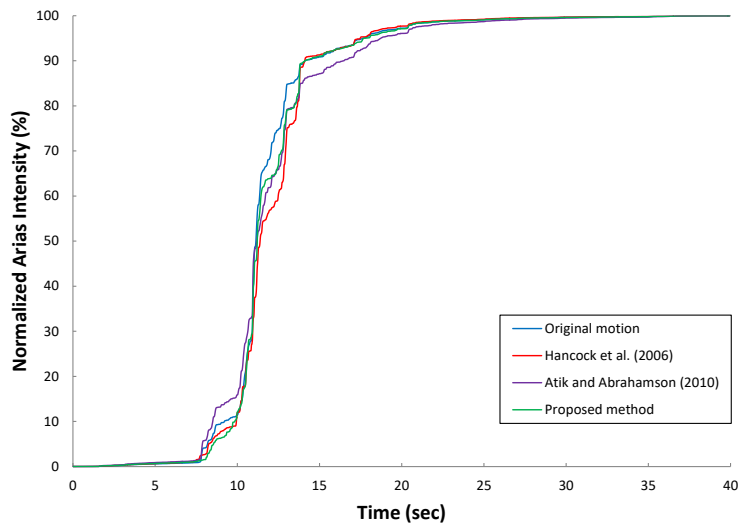

**Figure 14 Normalized Arias intensity of original accelerogram beside the adjusted motions using different spectral matching techniques**

## 6. Summary and conclusion

An innovative method—based on optimization and discrete wavelet transformation (DWT)—is introduced for the spectral matching procedure. When this optimization-based spectral matching is applied to an initial ground motion, the motion would be adjusted in such that its response spectrum is matched to a pre-selected target spectrum. While optimization methods are potent tools for solving complicated problems, they are quite straightforward to be implemented. Through using a nonlinear least-squares optimization method, the DWT coefficients associated with a signal of an initial motion are adjusted in a way that its related acceleration response spectrum gets fitted to the target spectrum. By adding two sine functions in this framework, a condition to impose zero value in the displacement and velocity time history at the end of a motion is satisfied. Furthermore, in the formulation of the proposed method, multiple damping levels can be simply defined for the objective function. In this case, the proposed approach is applied to an initial motion against the targets which are defined with four damping ratios—i.e., 5%, 10%, 20%, and 30%. Any desired constraint can be generally added to the optimization objective function of this proposed matching framework. Although it has not been applied to the presented example, decision variables can be found through an imposed constraint by which amplitudes of adjusted time histories are kept very close to their equivalent counterparts of the original motion. In order to investigate the effectiveness of the proposed method, first it is applied to a case study and then compared with two existing spectral matching approaches. The results are as follows:

1. The results of the matching procedure at four different damping ratios show that the proposed algorithm can adjust a ground motion while the energy content of the motion changes only by 1% as compared to the original motion.

2. The non-stationary characteristics, as well as the frequency content of the initial time series, are preserved with negligible changes within the adjusted motion. Also, the energy content of the original motion is altered less than 10% after matching.

3. In the time series of the adjusted motion profiles, the velocity and displacement at the end of a motion get to zero value through primitive baseline correction. More, given the fact that the modified signals are also filtered by a standard baseline correction implemented in the proposed algorithm, there is no need to use a separate baseline correction for modified time series.

4. Spectrum-compatible time series generated by the proposed method clearly demonstrate an improved misfit of less than 0.5% once they are compared to adjusted motions matched by other conventional methods.